\documentclass[twocolumn,showpacs,preprintnumbers,amsmath,amssymb]{revtex4}
\usepackage{graphicx}
\begin{document}
\preprint{123XXXX}
\title{\textbf{Tsallis' }$q$\textbf{\ index and Mori's }$q$ \textbf{phase\
transitions at edge of chaos}}
\author{E. Mayoral and A. Robledo
\thanks{E-mail addresses:robledo@fisica.unam.mx }}
\address{Instituto de F\'{i}sica,\\
Universidad Nacional Aut\'{o}noma de M\'{e}xico,\\
Apartado Postal 20-364, M\'{e}xico 01000 D.F., Mexico.}
\date{2005}
\begin{abstract}

We uncover the basis for the validity of the Tsallis statistics at the onset
of chaos in logistic maps. The dynamics within the critical attractor is
found to consist of an infinite family of Mori's $q$-phase transitions of
rapidly decreasing strength, each associated to a discontinuity in
Feigenbaum's trajectory scaling function $\sigma $. The value of $q$ at each
transition corresponds to the same special value for the entropic index $q$,
such that the resultant sets of $q$-Lyapunov coefficients are equal to the
Tsallis rates of entropy evolution.

\end{abstract}

\pacs{05.90.+m, 0.5.10.Cc, 05.45.Ac}
\maketitle

\section{Introduction}
Searches for evidence of nonextensive \cite{tsallis0}, \cite{tsallis1}
properties at the period-doubling onset of chaos in logistic maps - the
Feigenbaum attractor - have at all times yielded affirmative responses, from
the initial numerical studies \cite{tsallis2}, to subsequent heuristic
investigations \cite{tsallis3}, and the more recent rigorous results \cite%
{robledo1}, \cite{robledo2}. However a critical analysis and a genuine
understanding of the basis for the validity at this attractor of the
nonextensive generalization \cite{tsallis0}, \cite{tsallis1} of the
Boltzmann-Gibbs (BG) statistical mechanics - here referred as $q$-statistics
- is until now lacking. Here we clarify the circumstances under which the
features of $q$-statistics are observed and, most importantly, we
demonstrate that the mechanism by means of which the Tsallis entropic index $%
q$ arises is provided by the occurrence of dynamical phase transitions of
the kind described by the formalism of Mori and colleagues \cite{mori1}.
These transitions, similar to first order thermal phase transitions, are
associated to trajectories that link different regions within a multifractal
attractor. The onset of chaos is an incipiently chaotic attractor, with
memory preserving, nonmixing, phase space trajectories. Because many of its
properties are familiar, and well understood since many years ago, it is of
interest to explain how previous knowledge fits in with the new perspective.

The Feigenbaum attractor is the classic one-dimensional critical attractor
with universal properties in the renormalization group (RG) sense, i.e.
shared by all unimodal (one hump) maps with the same degree of nonlinearity.
The static or geometrical properties of this attractor are understood since
long ago \cite{schuster1} - \cite{hilborn1}, and are represented, for
example, by the generalized dimensions $D(\mathsf{q})$ or the spectrum of
dimensions $f(\widetilde{\alpha })$ that characterize the multifractal set 
\cite{beck1}, \cite{hilborn1}. The dynamical properties that involve
positions within the attractor also display universality and, as we see
below, these are conveniently given in terms of the discontinuities in
Feigenbaum's trajectory scaling function $\sigma $ that measures the
convergence of positions in the orbits of period $2^{n}$ as $n\rightarrow
\infty $ \cite{schuster1}. Let us first recall that the Feigenbaum attractor
has a vanishing ordinary Lyapunov coefficient $\lambda _{1}$ and that the
sensitivity to initial conditions $\xi _{t}$ does not converge to any
single-valued function and displays fluctuations that grow indefinitely \cite%
{grassberger1}, \cite{politi1}, \cite{mori2}, \cite{mori1}. For initial
positions at the attractor $\xi _{t}$ develops a universal self-similar
temporal structure and its envelope grows with $t$ as a power law \cite%
{grassberger1}, \cite{politi1}, \cite{mori2}, \cite{mori1}, \cite{tsallis2}.
We are interested here in determining the detailed dependence of the
aforementioned structure on \textit{both} the initial position $x_{0}$ and
the observation time $t$ as this dependence is preserved by the infinitely
lasting memory. Therefore we shall not consider the effect of averaging with
respect to $x_{0}$ and/or $t$, explored in other studies \cite{grassberger1} 
\cite{tsallis4}, as this would obscure the fine points of the dynamics.

The central assertion of the $q$-statistics with regards to the dynamics of
critical attractors is a sensitivity to initial conditions $\xi _{t}$
associated to the $q$-exponential functional form, i.e. the '$q$-deformed'
exponential function $\exp _{q}(x)\equiv [1-(q-1)x]^{-1/(q-1)}$. From such $%
\xi _{t}$ a $q$-generalized Lyapunov coefficient $\lambda _{q}$ can be
determined just as $\lambda _{1}$ is read from an exponential $\xi _{t}$.
The $\lambda _{q}$ is presumed to satisfy a $q$-generalized identity $%
\lambda _{q}=K_{q}$ \cite{note1} \cite{hilborn1} where $K_{q}$ is an entropy
production rate based on the Tsallis entropy $S_{q}$, defined in terms of
the $q$-logarithmic function $\ln _{q}y\equiv (y^{1-q}-1)/(1-q)$, the
inverse of $\exp _{q}(x)$. Unlike $\lambda _{1}$ for (ergodic) chaotic
attractors, the coefficient $\lambda _{q}$ is dependent on the initial
position $x_{0}$ and therefore $\lambda _{q}$ constitutes a spectrum (and
also $K_{q}$) that can be examined by varying this position.

The \emph{fixed} values of the entropic index $q$ are obtained from the
universality class parameters to which the attractor belongs. For the
simpler pitchfork and tangent bifurcations there is a single well-defined
value for the index $q$ for each type of attractor as a single $q$%
-exponential describes the sensitivity \cite{robledo4}. For multifractal
critical attractors the situation is more complicated and there appear to be
a multiplicity of indexes $q$ but with precise values given by the attractor
scaling functions. As shown below, the sensitivity takes the form of a
family of interweaved $q$-exponentials. The $q$-indexes appear in conjugate
pairs, $q$ and $Q=2-q$, as these correspond to switching starting and
finishing trajectory positions. We show that $q$ and $Q$ are related to the
occurrence of pairs of dynamical '$q$-phase' transitions that connect
qualitatively different regions of the attractor \cite{mori1} \cite{mori2}.
These transitions are identified as the source of the special values for the
entropic index $q$. For the Feigenbaum attractor an infinite family of such
transitions take place but of rapidly decreasing strength.

In the following section we recall the essential features of the
statistical-mechanical formalism of Mori and colleagues \cite{mori1} to
study dynamical phase transitions in attractors of nonlinear maps and follow
this by a summary of expressions of the $q$-statistics. Then, in subsequent
sections we present known properties and develop others for the dynamics
within the Feigenbaum attractor. Amongst these we derive the sensitivity $%
\xi _{t}$ in terms of the trajectory scaling function $\sigma $, and use
this to make contact with both Mori's and Tsallis' schemes. We discuss our
results.

\section{Statistical mechanics for critical attractors}

During the late 1980's Mori and coworkers developed a comprehensive
thermodynamic formalism to characterize drastic changes at bifurcations and
at other singular phenomena in low dimensional maps \cite{mori1}. The
formalism was further adapted to study critical attractors and was
illustrated by considering the specific case of the period-doubling onset of
chaos in the logistic map \cite{politi1}, \cite{mori2}, \cite{mori1}. For
critical attactors the scheme involves the evaluation of fluctuations of the
generalized finite-time Lyapunov coefficient 
\begin{equation}
\lambda (t,x_{0})=\frac{1}{\ln t}\sum_{i=0}^{t-1}\ln \left| \frac{df_{\mu
_{\infty }}(x_{i})}{dx_{i}}\right| ,\quad t\gg 1,  \label{lyapunovdef}
\end{equation}
where $f_{\mu }(x)$ is here the logistic map, or its extension to
non-linearity of order $z>1$, 
\begin{equation}
f_{\mu }(x)=1-\mu \left| x\right| ^{z},\;-1\leq x\leq 1,\;0\leq \mu \leq 2.
\label{z-logistic1}
\end{equation}
We denote by $\mu _{\infty }(z)$ the value of the control parameter $\mu $
at the onset of chaos, with $\mu _{\infty }(2)=1.40115...$. Notice the
replacement of the customary $t$ by $\ln t$ in Eq. (\ref{lyapunovdef}), as
the ordinary Lyapunov coefficient $\lambda _{1}$ vanishes for critical
attractors, here at $\mu _{\infty }$, $t\rightarrow \infty $.

The density distribution for the values of $\lambda $, at $t\gg 1$, $%
P(\lambda ,t)$, is written in the form \cite{mori1} \cite{mori2} 
\begin{equation}
P(\lambda ,t)=t^{-\psi (\lambda )}P(0,t),  \label{DistLyapMori1}
\end{equation}
where $\psi (\lambda )$ is a concave spectrum of the fluctuations of $%
\lambda $ with minimum $\psi (0)=0$ and is obtained as the Legendre
transform of the 'free energy' function $\phi (\mathsf{q})$, defined as 
\begin{equation}
\phi (\mathsf{q})\equiv -\lim_{t\rightarrow \infty }\frac{\ln Z(t,\mathsf{q})%
}{\ln t},  \label{freeMori1}
\end{equation}
where $Z(t,\mathsf{q})$ is the dynamic partition function 
\begin{equation}
Z(t,\mathsf{q})\equiv \int d\lambda \ P(\lambda ,t)\ t^{-(\mathsf{q}%
-1)\lambda }.  \label{PartMori1}
\end{equation}
The 'coarse-grained' function of generalized Lyapunov coefficients $\lambda (%
\mathsf{q})$ and the variance $v(\mathsf{q})$ of $P(\lambda ,t)$ are given,
respectively, by 
\begin{equation}
\lambda (\mathsf{q})\equiv \frac{d\phi (\mathsf{q})}{d\mathsf{q}}\text{ and }%
v(\mathsf{q})\equiv \frac{d\lambda (\mathsf{q})}{d\mathsf{q}}
\label{lambdaandnu1}
\end{equation}
\cite{mori1} \cite{mori2}. Notice the special weight $t^{-(\mathsf{q}%
-1)\lambda }$ in $Z(t,\mathsf{q})$ and in the quantities derived from it.
The functions $\phi (\mathsf{q})$ and $\psi (\lambda )$ are the dynamic
counterparts of the Renyi dimensions $D(\mathsf{q})$ and the spectrum $f(%
\widetilde{\alpha })$ that characterize the geometric structure of the
attractor \cite{beck1}.

As with ordinary thermal 1st order phase transitions, a ''$q$-phase''
transition is indicated by a section of linear slope $m_{c}=1-q$ in the
spectrum (free energy) $\psi (\lambda )$, a discontinuity at $\mathsf{q}=q$
in the Lyapunov function (order parameter) $\lambda (\mathsf{q})$, and a
divergence at $q$ in the variance (susceptibility) $v(\mathsf{q})$. For the
onset of chaos at $\mu _{\infty }(z=2)$ a single $q$-phase transition was
numerically determined \cite{mori1} - \cite{politi1} and found to occur at a
value close to $m_{c}=-(1-q)\simeq -0.7$. Arguments were provided \cite%
{mori1} - \cite{politi1} for this value to be $m_{c}=-(1-q)=-\ln 2/\ln
\alpha =-0.7555...$, where $\alpha =2.50290...$ is one of the Feigenbaum's
universal constants. Our analysis below shows that this initial result gives
a broad picture of the dynamics at the Feigenbaum attractor and that
actually an infinite family of $q$-phase transitions of decreasing magnitude
takes place at $\mu _{\infty }$.

Independently, Tsallis and colleagues proposed \cite{tsallis2} that for
critical attractors the sensitivity to initial conditions $\xi _{t}$
(defined as $\xi _{t}(x_{0})\equiv \lim_{\Delta x_{0}\to 0}(\Delta
x_{t}/\Delta x_{0})$ where $\Delta x_{0}$ is the initial separation of two
orbits and $\Delta x_{t}$ that at time $t$), has the form 
\begin{equation}
\xi _{t}(x_{0})=\exp _{q}[\lambda _{q}(x_{0})\ t],  \label{sensitivity1}
\end{equation}
that yields the customary exponential $\xi _{t}$ with $\lambda _{1}$ when $%
q\rightarrow 1$. In Eq. (\ref{sensitivity1}) $q$ is the entropic index and
the initial-position-dependent $\lambda _{q}(x_{0})$ are the $q$-generalized
Lyapunov coefficients. Tsallis and colleagues also suggested \cite{tsallis2}
that the identity $K_{1}=$ $\lambda _{1}$ \cite{note1}, where the rate of
entropy production $K_{1}$ is given by 
\begin{equation}
K_{1}t=S_{1}(t)-S_{1}(0),\;t\;\text{large},
\end{equation}
and 
\begin{equation}
S_{1}=-\sum_{i}p_{i}\ln p_{i},  \label{BGentropy1}
\end{equation}
for an ensemble of trajectories with instantaneous distribution $p_{i}$,
would be generalized to $K_{q}=$ $\lambda _{q}$, where the $q$-generalized
rate of entropy production $K_{q}$ is defined via 
\begin{equation}
K_{q}t=S_{q}(t)-S_{q}(0),\;t\;\text{large},
\end{equation}
and where 
\begin{equation}
S_{q}\equiv \sum_{i}p_{i}\ln _{q}p_{i}^{-1}=\frac{1-\sum_{i}^{W}p_{i}^{q}}{%
q-1}  \label{tsallisentropy1}
\end{equation}
is the Tsallis entropy. These properties have been corroborated to hold at $%
\mu _{\infty }(z)$ numerically \cite{tsallis2}, \cite{tsallis4} for sets of
trajectories with $x_{0}$ spread throughout $-1\leq x_{0}\leq 1$ and
analytically for specific classes of trajectories starting near $x_{0}=0$
and observed at specific times of the form $t=(2k+1)2^{n}-1$, $k=0,1,2...$
and $n=1,2,...$ \cite{robledo1} - \cite{robledo3}. We explain the rationale
for these particular choices of $x_{0}$ and $t$ in the following section
where we examine the structure of trajectories inside the attractor. See
Fig. 1.

\begin{figure*}[tbp]
\includegraphics[width=6cm ,angle=-90]{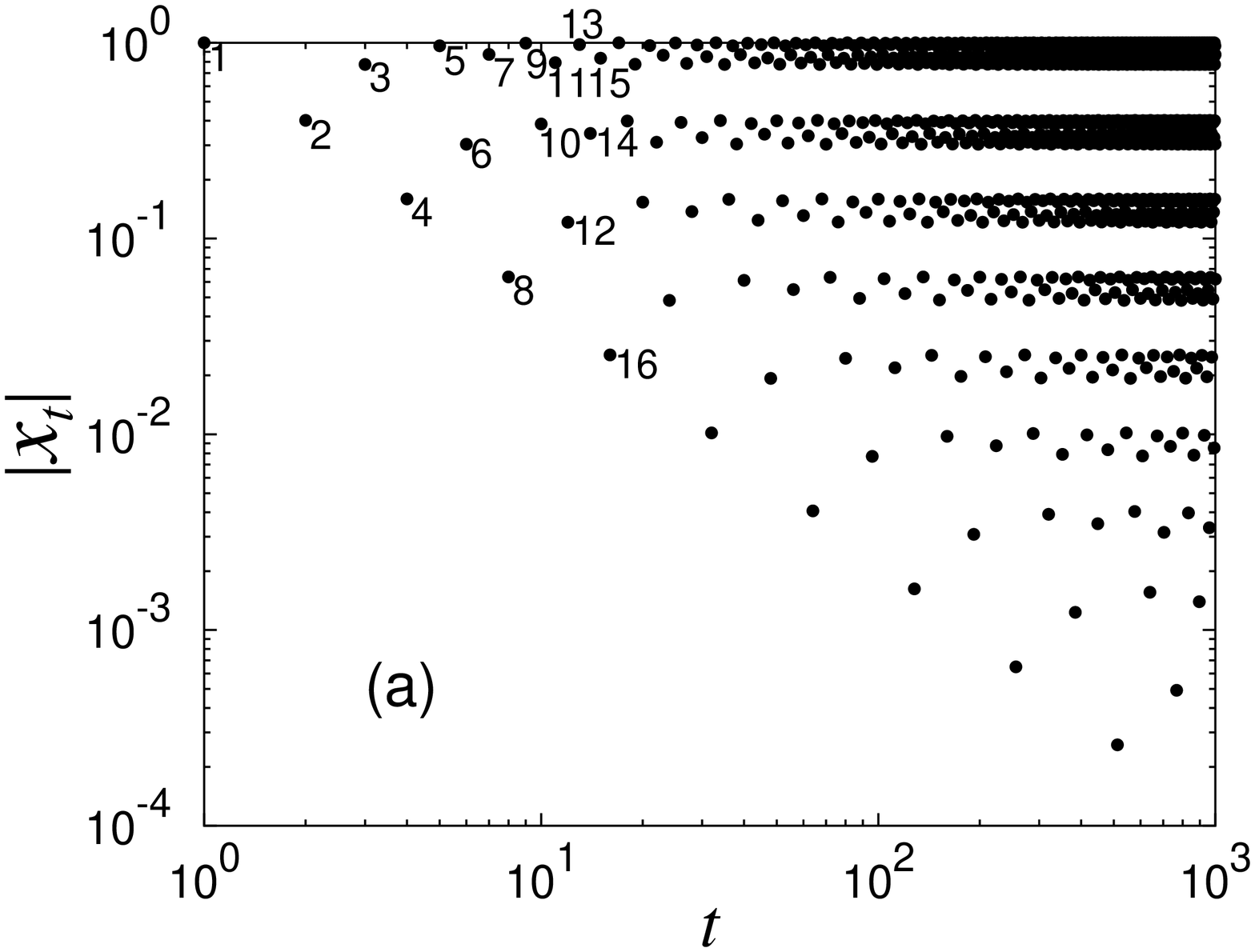}
\includegraphics[width=6cm ,angle=-90]{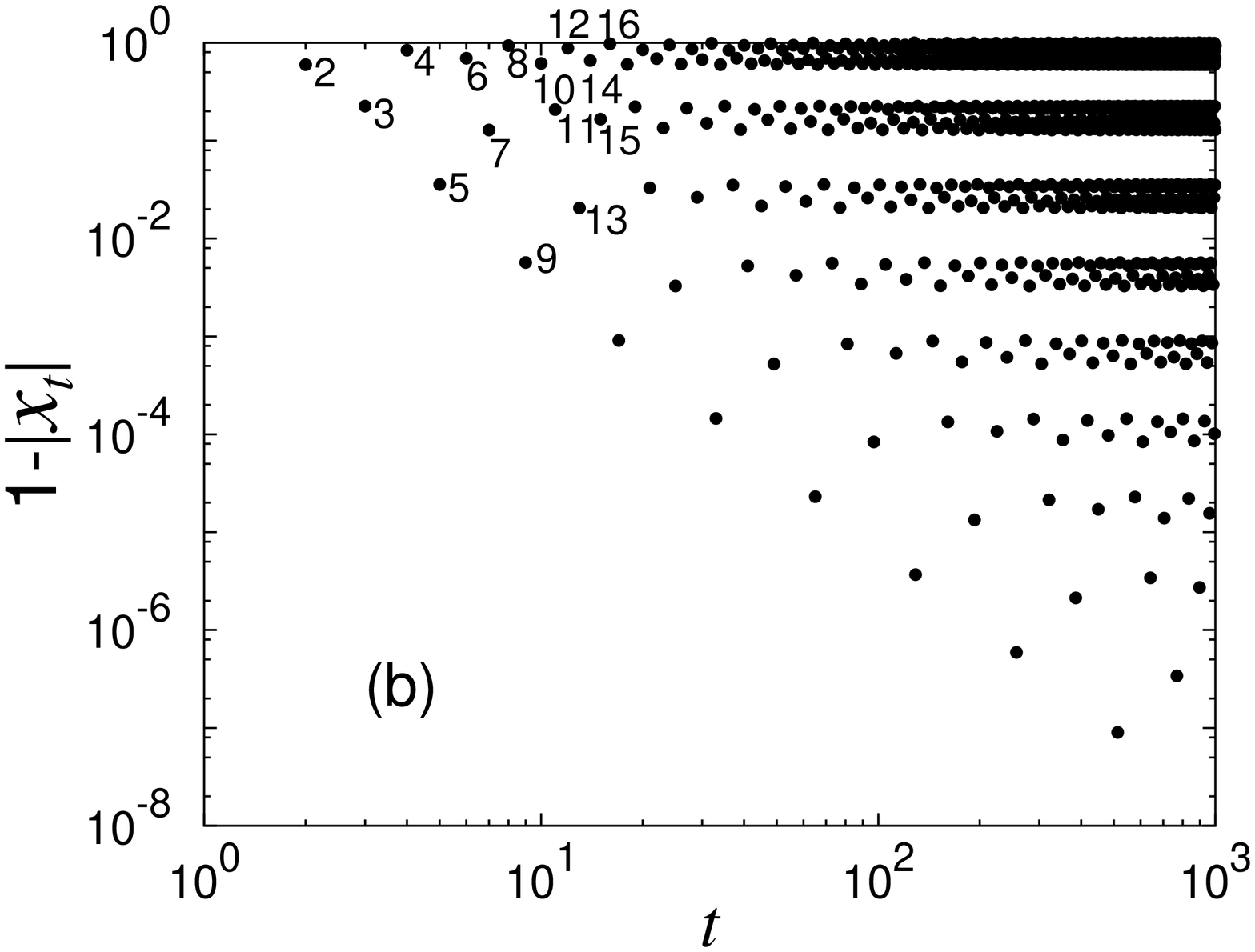}
\caption{{\protect\small a) Absolute values of }$\left\vert x_{t}\right\vert 
${\protect\small \ vs }$t${\protect\small \ in logarithmic scales for the
orbit with initial condition }$x_{0}=0${\protect\small \ at }$\protect\mu %
_{\infty }${\protect\small \ of the logistic map }$z=2${\protect\small . The
labels indicate iteration time }$t${\protect\small . b) Same as a) with }$%
\left\vert x_{t}\right\vert ${\protect\small \ replaced by }$1-\left\vert
x_{t}\right\vert ${\protect\small .}}
\label{fig1}
\end{figure*}

\section{Dynamics within the Feigenbaum attractor}

In Fig. 1a we have plotted (in logarithmic scales) the first few absolute
values of iterated positions $\left| x_{t}\right| $ of the orbit at $\mu
_{\infty }(z=2)$ starting at $x_{0}=0$ where the labels indicate iteration
time $t$. Notice the structure of horizontal bands, and that in the top band
lie half of the attractor positions (odd times), in the second band a
quarter of the attractor positions, and so on. The top band is eliminated by
functional composition of the original map, that is by considering the orbit
generated by the map $f_{\mu _{\infty }}^{(2)}(0)$ instead of $f_{\mu
_{\infty }}(0)$. Successive bands are eliminated by considering the orbits
of $f_{\mu _{\infty }}^{(2^{j})}(0),\ j=1,2...$. The positions of the top
band (odd times) can be reproduced approximately by the positions of the
band below it (times of the form $t=2+4n$, $n=0,1,2...$) by multiplication
by a factor equal to $\alpha $, e.g. $\left| x_{1}\right| \simeq \alpha
\left| x_{2}\right| $. Likewise, the positions of the second band are
reproduced by the positions of the third band under multiplication by $%
\alpha $, e.g. $\left| x_{2}\right| \simeq \alpha \left| x_{4}\right| $, and
so on. In Fig. 1b we show a logarithmic scale plot of $1-$ $\left|
x_{t}\right| $ that displays a band structure similar to that in Fig. 1a, in
the top band lie again half of the attractor positions (even times) and
below the other half (odd times) is distributed in the subsequent bands.
This time the positions in one band are reproduced approximately from the
positions of the band lying below it by multiplication by a factor $\alpha
^{z}$, e.g. $1-\left| x_{3}\right| \simeq \alpha ^{z}(1-\left| x_{5}\right|
) $. The properties amongst bands of iterate positions merely follow from
repeated composition and rescaling of the map and represent a graphical
construction of the Feigenbaum RG transformation $Rf(x)\equiv \alpha
f(f(x/\alpha ))$, where $\alpha (z=2)=2.50290...$

Also, the trajectory with initial condition $x_{0}=0$ maps out the
Feigenbaum attractor in such a way that the absolute values of succeeding
(time-shifted $\tau =t+1$) positions $\left| x_{\tau }\right| $ form
subsequences with a common power-law decay of the form $\tau ^{-1/1-q}$ with 
$q=1-\ln 2/\ln \alpha (z)$, with $q\simeq 0.24449$ when $z=2$. See how
positions fall along straight diagonal lines in Fig. 1a. That is, the 
\textit{entire} attractor can be decomposed into position subsequences
generated by the time subsequences $\tau =(2k+1)2^{n}$, each obtained by
running over $n=0,1,2,...$ for a fixed value of $k=0,1,2,...$ Noticeably,
the positions in these subsequences can be obtained from those belonging to
the 'superstable' periodic orbits of lengths $2^{n}$, i.e. the $2^{n}$%
-cycles that contain the point $x=0$ at $\overline{\mu }_{n}<\mu _{c}(0)$ 
\cite{schuster1}. Specifically, the positions for the main subsequence $k=0$%
, that constitutes the lower bound of the entire trajectory (see Fig. 1a),
can be identified to be $\left| x_{2^{n}}\right| \simeq d_{n,0}=$ $\alpha
^{-n}$, where $d_{n,0}\equiv \left| f_{\overline{\mu }_{n}}^{(2^{n-1})}(0)%
\right| $ is the '$n$-th principal diameter' defined at the $2^{n}$%
-supercycle, the distance of the orbit position nearest to $x=0$ \cite%
{schuster1}. The main subsequence can be expressed as 
\begin{equation}
\left| x_{t}\right| =\exp _{2-q}(-\Lambda _{q}t)
\end{equation}
with $\Lambda _{q}=(z-1)\ln \alpha /\ln 2$. Interestingly this analytical
result for $\left| x_{t}\right| $ can be seen to satisfy the dynamical
fixed-point relation, $h(t)=\alpha h(h(t/\alpha ))$ with $\alpha
=2^{1/(1-q)} $. See \cite{robledo1}, \cite{robledo2} for $z=2$ and \cite%
{robledo3} for general $z>1$.

We now work out the relationship between the trajectory scaling function $%
\sigma $ and the sensitivity $\xi _{t}$ at $\mu _{\infty }$. To begin with
we recall \cite{schuster1} the general definition of the diameters $d_{n,m}$
that measure the bifurcation forks that form the period-doubling cascade
sequence. The $d_{n,m}$ in these orbits are defined as the distances of the
elements $x_{m}$, $m=0,1,2,...,2^{n}-1$, to their nearest neighbors $f_{%
\overline{\mu }_{n}}^{(2^{n-1})}(x_{m})$, i.e. 
\begin{equation}
d_{n,m}\equiv f_{\overline{\mu }_{n}}^{(m+2^{n-1})}(0)-f_{\overline{\mu }%
_{n}}^{(m)}(0).  \label{diameter1}
\end{equation}
For large $n$, $d_{n,0}/d_{n+1,0}\simeq -\alpha (z)$; $\alpha (2)=\alpha $.
Further, Feigenbaum \cite{feigenbaum1} constructed the auxiliary function 
\begin{equation}
\sigma _{n}(m)=\frac{d_{n+1,m}}{d_{n,m}}  \label{sigma1}
\end{equation}
to quantify the rate of change of the diameters and showed that in the limit 
$n\rightarrow \infty $ it has finite (jump) discontinuities at all rationals
of the form $m/2^{n+1}$. So, considering the variable $y=m/2^{n+1}$ one
obtains \cite{feigenbaum1} \cite{schuster1}, omitting the subindex $n$, $%
\sigma (0)=-1/\alpha $, but $\sigma (0^{+})=1/\alpha ^{z}$, and through the
antiperiodic property $\sigma (y+1/2)=-\sigma (y)$, also $\sigma
(1/2)=1/\alpha $, but $\sigma (1/2+0^{+})=-1/\alpha ^{z}$. Other
discontinuities in $\sigma (y)$ appear at $y=1/4,1/8,3/8$,$\ $etc. In most
cases it is only necessary to consider the first few as their magnitude
decreases rapidly. See, e.g. Fig. 31 in Ref. \cite{schuster1}. The
discontinuities of $\sigma _{n}(m)$ can be suitably obtained by first
generating the superstable orbit $2^{\infty }$ at $\mu _{\infty }$ and then
plotting the position differences $\left| x_{t}-x_{m}\right| =\left| f_{\mu
_{\infty }}^{(t)}(0)-f_{\mu _{\infty }}^{(m)}(0)\right| $ for times of the
form $t=2^{n}-m$, $n=0,1,2,...$, in logarithmic scales. The distances that
separate positions along the time subsequence correspond to the logarithm of
the diameters $d_{n,m}$. See Figs. 1a and 1b where the constant spacing of
positions along the main diagonal provide the values for $\ln d_{n,0}\simeq
-n\ln \alpha $ and $\ln d_{n,1}\simeq -n\ln \alpha ^{z}$, respectively. The
constant slope $s_{m}$ of the resulting time subsequence data is related to $%
\sigma _{n}(m)$, i.e. $\sigma _{n}(m)=2^{s_{m}}$. See Figs. 1a and 1b where
the slopes of the main diagonal subsequences have the values $-\ln \alpha
/\ln 2$ and $-z\ln \alpha /\ln 2$, respectively. From these two slopes the
value of the largest jump discontinuity of $\sigma _{n}(m)$ is conviniently
determined.

A key factor in obtaining our results is the fact that the sensitivity $\xi
_{t}(x_{0})$ can be evaluated for trajectories within the attractor via
consideration of the discontinuities of $\sigma (y)$. Our strategy for
determining $\xi _{t}(x_{0})$ is to chose the initial and the final
separation of the trajectories to be the diameters $\Delta x_{0}=d_{n,m}$
and $\Delta x_{t}=d_{n,m+t}$, $t=2^{n}-1$, respectively. Then, $\xi
_{t}(x_{0})$ is obtained as 
\begin{equation}
\xi _{t}(x_{0})=\lim_{n\rightarrow \infty }\left| \frac{d_{n,m+t}}{d_{n,m}}%
\right| ^{n}.
\end{equation}
Notice that in this limit $\Delta x_{0}\rightarrow 0$, $t\rightarrow \infty $
\textit{and} the $2^{n}$-supercycle becomes the onset of chaos (the $%
2^{\infty }$-supercycle). Then, $\xi _{t}(x_{0})$ can be written as 
\begin{equation}
\xi _{t}(m)\simeq \left| \frac{\sigma _{n}(m-1)}{\sigma _{n}(m)}\right|
^{n},\ t=2^{n}-1,\;n\;\text{large},  \label{sensitivity2}
\end{equation}
where we have used $\left| \sigma _{n}(m)\right| ^{n}\simeq
\prod_{i=1}^{n}\left| d_{i+1,m}/d_{i,m}\right| $ and $%
d_{i+1,m+2^{n}}=-d_{i+1,m}$. Notice that for the inverse process, starting
at $\Delta x_{0}=d_{n,m+t}=-d_{n,m-1}$ and ending at $\Delta x_{t^{\prime
}}=d_{n,m}=-d_{n,m-1+t^{\prime }}$, with $t^{\prime }=2^{n}+1$ one obtains 
\begin{equation}
\xi _{t^{\prime }}(m-1)\simeq \left| \frac{\sigma _{n}(m)}{\sigma _{n}(m-1)}%
\right| ^{n},\ t^{\prime }=2^{n}+1,\;n\;\text{large}.
\end{equation}

Given the known properties of $\sigma _{n}(m)$ we can readily extract those
for $\xi _{t}(m)$ for general non-linearity $z>1$. Taking into account only
the first $2M$, $M=1,2,...$ discontinuities of $\sigma _{n}(m)$ we have the
antiperiodic step function 
\begin{equation}
1/\sigma _{n}(m)=\left\{ 
\begin{array}{l}
\alpha _{l},\;l2^{n-M}\leq m<(l+1)2^{n-M} \\ 
-\alpha _{l},\;(2^{M}+l)2^{n-M}\leq m<(2^{M}+l+1)2^{n-M},%
\end{array}
\right.
\end{equation}
with $l=0,1,...,2^{M}-1$, and this implies that $\xi _{t}(m)=1$ when $\sigma
_{n}(m)$ is continuous at $m$, and 
\begin{equation}
\xi _{t}(m)=\left| \frac{\alpha _{l}}{\alpha _{l+1}}\right| ^{n},\;\text{or}%
,\;\;\xi _{t^{\prime }}(m)=\left| \frac{\alpha _{l}}{\alpha _{l+1}}\right|
^{-n},  \label{sensitivity?}
\end{equation}
when $\sigma _{n}(m)$ has a discontinuity at $m=l2^{n-M}$. As we make clear
below, these behaviors for the sensitivity reflect the multi-region nature
of the multifractal attractor and the memory retention of these regions in
the dynamics. Thus, $\xi _{t}(m)=1$ (or $\lambda _{q}(x_{0})=0$) corresponds
to trajectories that depart and arrive in the same region, while the power
laws in Eq. (\ref{sensitivity?}) correspond to a departing position in one
region and arrival at a different region and vice versa, the trajectories
expand in one sense and contract in the other.

\section{Origin of Tsallis' $q$ index}

We now make explicit the mentioned link between the occurrence of Mori's $q$%
-phase transitions and the $q$-statistical dynamical properties at $\mu
_{\infty }$. Consider $M=1$, the simplest approximation to $\sigma _{n}(m)$,
yet it captures the effect on $\xi _{t}$ of the most dominant trajectories
within the attractor. This is to assume that half of the diameters scales as 
$\alpha _{0}=\alpha ^{z}$ (as in the most crowded region of the attractor, $%
x\simeq 1$) while the other half scales as $\alpha _{1}=\alpha $ (as in the
most sparse region of the attractor, $x\simeq 0$). With use of the identity $%
A^{n}=$ $(1+t/(2k+1))^{\ln A/\ln 2}$, $t=2^{n}-(2k+1)$, $\xi _{t}(m)=\left|
\alpha _{0}/\alpha _{1}\right| ^{n}$ can be rewritten as the $q$-exponential 
\begin{equation}
\xi _{t}(m)=\exp _{q_{0}}[\lambda _{q_{0}}^{(k)}\ t]\ ,
\label{sensitivity3a}
\end{equation}
where 
\begin{equation}
q_{0}=1-\frac{\ln 2}{(z-1)\ln \alpha },  \label{q_0index}
\end{equation}
and where 
\begin{equation}
\lambda _{q_{0}}^{(k)}=\frac{(z-1)\ln \alpha }{(2k+1)\ln 2},\;k=0,1,...
\label{lambdaq_0}
\end{equation}
The $(2k+1)^{-1}$ term in $\lambda _{q_{0}}^{(k)}$ arises from the time
shift involved in selecting different initial positions $x_{0}\simeq 1$. See
Fig. l and Ref. \cite{robledo2}. Similarly, the sensitivity for the
trajectories in the inverse order yields 
\begin{equation}
\xi _{t^{\prime }}(m)=\exp _{Q_{0}}[\lambda _{Q_{0}}^{(k)}\ t]\ ,
\label{sensitivity3b}
\end{equation}
where $Q_{0}=2-q_{0}$, and where $\lambda _{Q_{0}}^{(k)}=-2\lambda
_{q_{0}}^{(k)}$. The factor of $2$ in $\lambda _{Q_{0}}^{(k)}$ appears
because of a basic difference between the orbits of periods $2^{n}$ and $%
2^{\infty }$. In the latter case, to reach $x_{t^{\prime }}\simeq 0$ from $%
x_{0}\simeq 1$ at times $t^{\prime }=2^{n}+1$ the iterate necessarily moves
into positions of the next period $2^{n+1}$, and orbit contraction is twice
as effective than expansion. Notice that the relationship between the
indexes $Q_{0}=2-q_{0}$ for the couple of conjugate trajectories stems from
the property $\exp _{q}(x)=1/\exp _{2-q}(-x)$. For $z=2$ one obtains $%
Q_{0}\simeq 1.7555$ and $q_{0}\simeq 0.2445$, this latter value agrees with
that obtained in several earlier studies \cite{tsallis2} - \cite{robledo2}.
From the results for $\lambda _{q_{0}}^{(k)}$ and $\lambda _{Q_{0}}^{(k)}$
we can construct the two-step Lyapunov function 
\begin{equation}
\lambda (\mathsf{q})=\left\{ 
\begin{array}{l}
\lambda _{q_{0}}^{(0)},\;-\infty <\mathsf{q}\leq q_{0}, \\ 
0,\;\;\;\;\;q_{0}<\mathsf{q}<Q_{0}, \\ 
\lambda _{Q_{0}}^{(0)},\;\;\;Q_{0}\leq \mathsf{q}<\infty .%
\end{array}
\right.
\end{equation}
For $z=2$ one has $\lambda _{q_{0}}^{(0)}=\ln \alpha /\ln 2\simeq 1.323$ and 
$\lambda _{Q_{0}}^{(0)}=-2\lambda _{q_{0}}^{(0)}\simeq -2.646$. See Fig. 2a.

\begin{figure*}[tbp]
\includegraphics[width=5cm ,angle=-90]{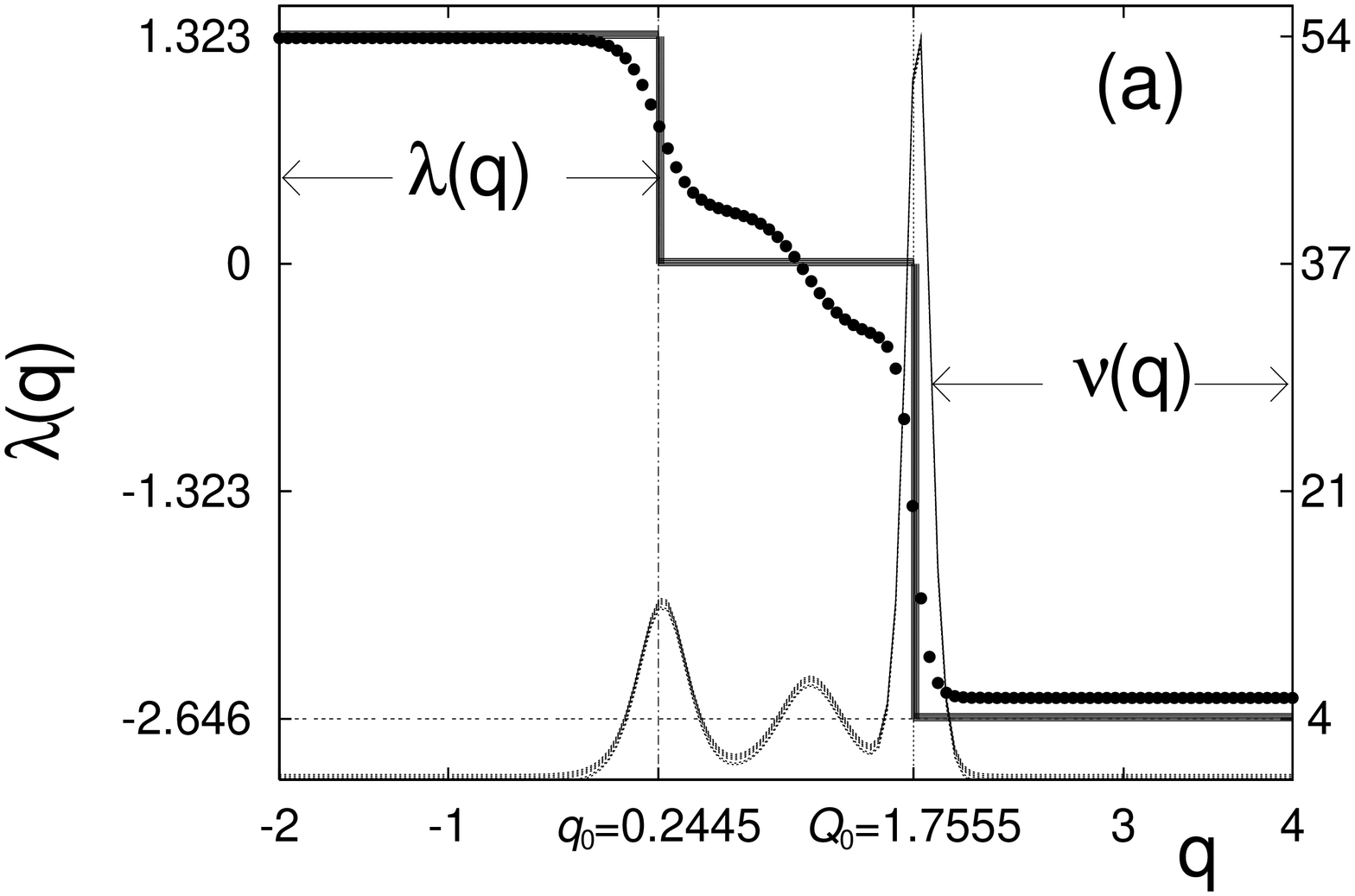} %
\includegraphics[width=5cm ,angle=-90]{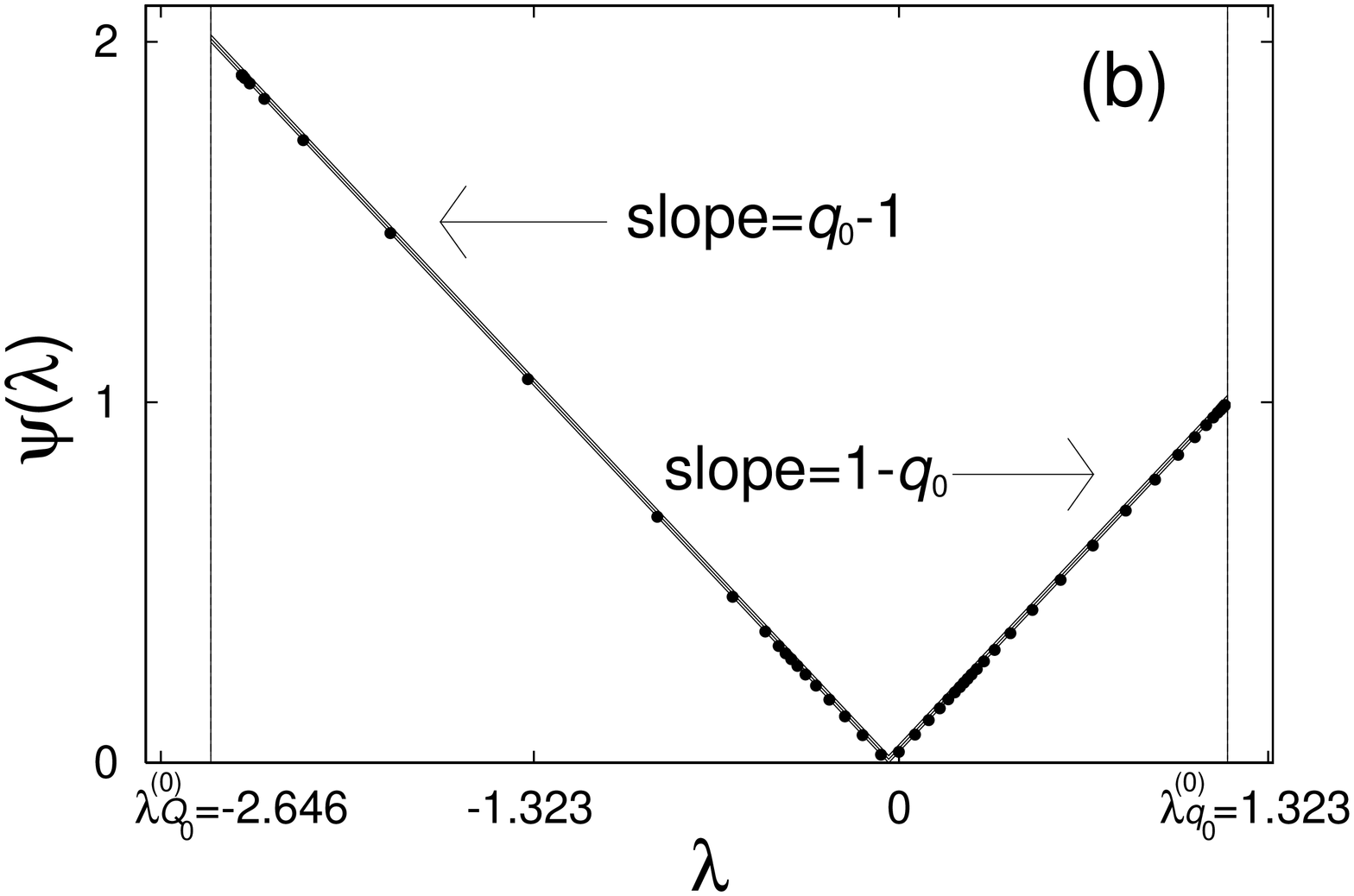}
\caption{$q${\protect\small -phase transitions for the main discontinuity in 
}$\protect\sigma _{n}(m)${\protect\small \ with index values }$q_{0}$%
{\protect\small \ and }$Q_{0}=2-q_{0}$ {\protect\small . The solid lines are
the piece-wise continuous functions given in the text while the solid
circles are the same functions calculated from the partition function }$%
Z(t,q)${\protect\small \ with the piece-wise }$\protect\psi (\protect\lambda %
)$ {\protect\small \ as input. See text for details.}}
\label{fig2}
\end{figure*}

When the next discontinuities of importance in $\sigma (y)$ are taken into
account new information on $\xi _{t}(x_{0})$ is obtained in the form of
additional pairs of $q$-exponentials as in Eqs. (\ref{sensitivity3a}) and (%
\ref{sensitivity3b}). In the next approximation for $\sigma _{n}(m)$ there
are four scaling factors, $\alpha _{0}=\alpha ^{z}$, $\alpha _{1}$, $\alpha
_{2}$, $\alpha _{3}=\alpha $, where $\alpha _{1}=1/\sigma (1/4)$ (with $%
\alpha _{1}\simeq 5.458$ for $z=2$) and $\alpha _{2}=1/\sigma (3/4)$ (with $%
\alpha _{2}\simeq 2.195$ for $z=2$). These last values are associated to the
two 'midway' regions between the most crowded and next most sparse regions
of the attractor. We have now three values for the $q$ index, $q_{0}$, $%
q_{1} $ and $q_{2}$ (together with the conjugate values $Q_{0}=2-q_{0}$, $%
Q_{1}=2-q_{1}$ and $Q_{2}=2-q_{2}$ for the inverse trajectories). For each
value of $q$ there is a set of $q$-Lyapunov coefficients running from a
maximum $\lambda _{q_{j},\max }$ to zero (or a minimum $\lambda _{Q_{j},\min
}$ to zero). From the results for $\lambda _{q_{j}}^{(k)}$ and $\lambda
_{Q_{j}}^{(k)}$, $j=0,1,2$ we can construct three Lyapunov functions $%
\lambda _{j}(\mathsf{q})$, $-\infty <\mathsf{q}<\infty $, each with two
jumps located at $\mathsf{q}=q_{j}=1-\ln 2/\ln \alpha _{j}(z)/\alpha (z)$
and $\mathsf{q}=Q_{j}=2-q_{j}$. Similar results are obtained when more
discontinuities in $\sigma _{n}(m)$ are taken into account.

Direct contact can be established now with the formalism developed by Mori
and coworkers and the $q$-phase transition reported in Ref. \cite{mori2}.
Each step function for $\lambda (\mathsf{q})$ can be integrated to obtain
the spectrum $\phi (\mathsf{q})$ ($\lambda (\mathsf{q})\equiv d\phi /d%
\mathsf{q}$) and from this its Legendre transform $\psi (\lambda )$ ($\equiv
\phi -(1-\mathsf{q})\lambda $). We illustrate this with the $\sigma _{n}(m)$
approximated with only two scale factors and present specific values when $%
z=2$. The free energy functions $\phi (\mathsf{q})$ and $\psi (\lambda )$
obtained from the two-step $\lambda (\mathsf{q})$ determined above and shown
in Fig. 2a are given by 
\begin{equation}
\phi (\mathsf{q})=\left\{ 
\begin{array}{l}
\lambda _{q_{0}}^{(0)}(\mathsf{q}-q_{0}),\;\;\mathsf{q}\leq q_{0}, \\ 
0,\;\;\;\;\;\;q_{0}<\mathsf{q}<Q_{0}, \\ 
\lambda _{Q_{0}}^{(0)}(\mathsf{q}-Q_{0}),\;\mathsf{q}\geq Q_{0},%
\end{array}
\right.
\end{equation}
and 
\begin{equation}
\psi (\lambda )=\left\{ 
\begin{array}{l}
(1-Q_{0})\lambda ,\;\lambda _{Q_{0}}^{(0)}<\lambda <0, \\ 
(1-q_{0})\lambda ,\;0<\lambda <\lambda _{q_{0}}^{(0)}.%
\end{array}
\right.
\end{equation}
See Fig. 2b. The constant slopes of $\psi (\lambda )$ represent the $q$%
-phase transitions associated to trajectories linking two regions of the
attractor, $x\simeq 1$ and $x\simeq 0$, and their values $1-q_{0}$ and $%
q_{0}-1$ correspond the index $q_{0}$ obtained for the $q$-exponentials $\xi
_{t}$ in Eqs. (\ref{sensitivity3a}) and (\ref{sensitivity3b}). The slope $%
q_{0}-1\simeq -0.7555$ coincides with that originally detected in Refs. \cite%
{mori2}, \cite{politi1}. When we consider also the next discontinuities of
importance in $\sigma (y)$ at $y=1/4$ , $3/4$ we obtain a couple of two $q$%
-phase transitions for each of the three values of the $q$ index, $q_{0}$, $%
q_{1}$ and $q_{2}$. The constant slope values for the $q$-phase transitions
at $1-q_{0}$ and $q_{0}-1$ appear again, but now we have two other pairs of
transitions with slope values $1-q_{1}$ and $q_{1}-1$, and, $1-q_{2}$ and $%
q_{2}-1$, that correspond, respectively, to orbits that link the most
crowded region of the attractor to the 'medium crowded' region, and to
orbits that link this 'medium crowded' region with the 'medium sparse'
region of the attractor.

\section{Equality between $q$-Lyapunov coefficients and rate of $q$-entropy
change}

Next, we verify the equality between the $q$-Lyapunov coefficients $\lambda
_{qi}^{(k)}$ and the $q$-generalized rates of entropy production $%
K_{q_{i}}^{(k)}$. We follow the same procedure as in Ref. \cite{robledo2}.
Consider a large number $\mathcal{N}$ of trajectories with initial positions
uniformly distributed within a small interval of length $\Delta x_{0}$
containing the attractor point $x_{0}$. A partition of this interval is made
with $N$ nonintersecting intervals of lengths $\varepsilon _{i,0}$ ($%
i=1,2,...,N$). For $\Delta x_{0}$ sufficiently small, after $t=2^{n}-1$, or $%
t^{\prime }=2^{n}+1$, iterations these interval lengths transform according
to 
\begin{equation}
\frac{\varepsilon _{i,t}}{\varepsilon _{i,0}}=\left| \frac{\alpha _{l}}{%
\alpha _{l+1}}\right| ^{n}\text{, or, }\frac{\varepsilon _{i,t^{\prime }}}{%
\varepsilon _{i,0}}=\left| \frac{\alpha _{l}}{\alpha _{l+1}}\right| ^{-n}.
\end{equation}
(Recall $\sigma _{n}(m)$ has a discontinuity at $m=l2^{n-M}$). We observe
that the interval ratios remain constant, that is, $\varepsilon
_{i,0}/\Delta x_{0}=\varepsilon _{i,t}/\Delta x_{t}$, since the
entire-interval ratio $\Delta x_{t}/\Delta x_{0}$ scales equally with $t$.
Thus, the initial number of trajectories within each interval $\mathcal{N}%
\varepsilon _{i,0}/\Delta x_{0}$ remains fixed in time, with the consequence
that the original distribution stays \emph{uniform} for all times $t<T$,
where $T\rightarrow \infty $ as $\Delta x_{0}\rightarrow 0$. We can now
calculate the rate of entropy production. This is more easily done with the
use of a partition of $W$ equal-sized cells of length $\varepsilon $. If we
denote by $W_{t}$ the number of cells that the ensemble occupies at time $t$
and by $\Delta x_{t}$ the total length of the interval$\ $these cells form,
we have $W_{t}=\Delta x_{t}/\varepsilon $ or $W_{t}=(\Delta x_{t}/\Delta
x_{0})(\Delta x_{0}/\varepsilon )$, and in the limit $\varepsilon
\rightarrow 0$, $\Delta x_{0}/\varepsilon \rightarrow 1$ we obtain the
simple result $W_{t}=$ $\xi _{t}$. As the distribution is uniform, and
recalling that Eq. (\ref{sensitivity1}) for $\xi _{t}$ holds in all cases,
the $q$-entropy is given by 
\begin{equation}
S_{q_{j}}(t)=\ln _{q_{j}}W_{t}=\lambda _{q_{j}}^{(k)}t,
\end{equation}
while $K_{q_{j}}^{(k)}=\lambda _{q_{j}}^{(k)}$, as $W_{t=0}=1$.

It is important to clarify the circumstances under which the equalities $%
\lambda _{q}^{(k)}=K_{q}^{(k)}$ are obtained as these could be interpreted
as shortcomings of the formalism. First of all, the rate $K_{q}$ does not
generalize the trajectory-based Kolmogorov-Sinai (KS) entropy $\mathcal{K}%
_{1}$ that is involved in the well-known Pesin identity $\lambda _{1}=%
\mathcal{K}_{1}$ \cite{schuster1} - \cite{hilborn1}. Presumably, the $q$%
-generalized KS entropy $\mathcal{K}_{q}$ would be defined in the same
manner as $\mathcal{K}_{1}$ with the use of $S_{q}$ in place of $S_{1}$. The
rate $K_{q}$ is determined from values of $S_{q}$ only at two different
times \cite{hilborn1}. The relationship between $\mathcal{K}_{1}$ and $K_{1}$
has been investigated for several chaotic maps \cite{latora1} and it has
been established that the equality $\mathcal{K}_{1}=K_{1}$ occurs during an
intermediate stage in the evolution of the entropy $S_{1}(t)$, after an
initial transient dependent on the initial distribution of positions and
before an asymptotic approach to a constant saturation value. Here we have
seemingly looked into the analogous intermediate regime in which one would
expect $\mathcal{K}_{q}=K_{q}$, as we explain below, however, the answer to
this question is not analyzed in this occasion.

We have only considered initial conditions within small distances outside
the positions of the Feigenbaum attractor and have not focused on the
initial transient behavior referred to in the above paragraph. As for the
final asymptotic regime mentioned above it should be kept in mind that the
distance between trajectories, from which we obtain $\lambda _{q}$, always
saturates because of the finiteness of the available phase space (the
multifractal subset of $[-1,1]$ that is the Feigenbaum attractor). It is
widely known \cite{hilborn1} that special care needs to be taken in
determining $\lambda _{1}$ from a time series to avoid saturation due to
folding and similar limitations occur for $\lambda _{q}^{(k)}$. Separation
of incipiently chaotic trajectories, just as separation of chaotic ones,
undergo two different processes, stretching which leads to the $q$%
-exponential regime in $\xi _{t}$ and folding which keeps the orbits
bounded. Therefore for $t$ sufficiently large Eq. (\ref{sensitivity1}) would
be no longer valid, just like the exponential $\xi _{t}$ of chaotic
attractors. This is the reason there is a saturation time $T$ in our
determination of $\lambda _{q}^{(k)}$ and this consequently supports our use
of the rates $K_{q}^{(k)}$ in $\lambda _{q}^{(k)}=K_{q}^{(k)}$.

\section{Concluding remarks}

Our most striking finding is that the dynamics at the onset of chaos is
constituted by an infinite family of Mori's $q$-phase transitions, each
associated to orbits that have common starting and finishing positions
located at specific regions of the attractor. Each of these transitions is
related to a discontinuity in the $\sigma $ function of 'diameter ratios',
and this in turn implies a $q$-exponential $\xi _{t}$ and a spectrum of $q$%
-Lyapunov coefficients - equal to the Tsallis rate of entropy production -
for each set of attractor regions. The transitions come in pairs with
specific conjugate indexes $q$ and $Q=2-q$, as these correspond to switching
starting and finishing orbital positions. Since the amplitude of the
discontinuities in $\sigma $ diminishes rapidly, in practical terms there is
only need of evaluation for the first few of them. The dominant
discontinuity is associated to the most crowded and sparse regions of the
attractor and this alone provides a very reasonable description of the
dynamics, as found in earlier studies \cite{tsallis2} - \cite{robledo2}. The
special values for the Tsallis entropic index $q$ in $\xi _{t}$ are equal to
the special values of the variable $q$ in the formalism of Mori and
colleagues at which the $q$-phase transitions take place. Therefore, we have
identified the cause or source for the entropic index $q$ observed at the
Feigenbaum attractor.

We found that the sensitivity to initial conditions at the onset of chaos
does not have the form of a single $q$-exponential but of infinitely many
interlaced $q$-exponentials. More precisely, we found a hierarchy of such
families of interlaced $q$-exponentials. An intricate state of affairs that
befits the rich scaling features of a multifractal attractor. This dynamical
organization is difficult to resolve from the consideration of a
straightforward time evolution, i.e. starting from an arbitrary position $%
x_{0}$ within the attractor and recorded at every time $t$. In this case
what is observed \cite{mori2} are strongly fluctuating quantities that
persist in time with a tangled pattern structure that presents memory
retention. On the other hand, if specific initial positions with known
location within the multifractal are chosen, and \ subsequent positions are
observed only at pre-selected times, when the trajectories visit another
region of choice, a well-defined $q$-exponential form for $\xi _{t}$
emerges. The specific value of $q$ and the associated Lyapunov\ spectrum $%
\lambda _{q}$ can be clearly determined. For each case the value of $q$ is
given by the values of the trajectory scaling function $\sigma $ at one of
its discontinuities, while the corresponding spectrum $\lambda _{q}$
reflects all starting positions in the multifractal region where the
trajectories originate. We remind the reader that the results presented here
are independent from the dynamics of approach to the attractor as we have
not considered the time evolution of initial positions $x_{0}$ outside the
attractor and leave this case for future attention.

Interestingly, the crossover from $q$-statistics to BG statistics can be
observed for control parameter values in the vicinity of the onset of chaos, 
$\mu \gtrsim \mu _{\infty }$, when the attractor consists of $2^{n}$ bands, $%
n$ large. The Lyapunov coefficient $\lambda _{1}$ of the chaotic attractor
decreases with $\Delta \mu =\mu -$ $\mu _{\infty }$ as $\lambda
_{1}\varpropto 2^{-n}\sim \Delta \mu ^{\kappa }$, $\kappa =\ln 2/\ln \delta
(z)$, where $\delta $ is the second Feigenbaum constant \cite{schuster1}.
The chaotic orbit consists of an interband periodic motion of period $2^{n}$
and an intraband chaotic motion. The expansion rate $\sum_{i=0}^{t-1}\ln
\left| df_{\mu }(x_{i})/dx_{i}\right| $ grows as $\ln t$ for $t<2^{n}$ but
as $t$ for $t\gg 2^{n}$ \cite{mori1}, \cite{mori2}. This translates as
Tsallis dynamics with $q\neq 1$ for $t<2^{n}$ but BG dynamics with $q=1$ for 
$t\gg 2^{n}$.

In summary, we have obtained further understanding about the nature of the
dynamics at the onset of chaos in logistic maps. We exhibited links between
original developments, such as Feigenbaum's $\sigma $ function and Mori's $q$%
-phase transitions, with more recent advances, such as $q$-exponential
sensitivity to initial conditions \cite{robledo1} and $q$-generalized
identity between Lyapunov coefficients and rate of entropy change \cite{robledo2}.
Chaotic orbits possess a time irreversible property that stems from mixing
in phase space and loss of memory, but orbits within critical attractors are
non-mixing and have no loss of memory. Our results apply to many other
unimodal one-parameter families of maps. One example is the exponential
function map $f(y)=1-\nu \exp (|y|^{-z})$ studied in Ref. \cite{tsallis4} as
the simple (monotonic) change of variable $|x|^{z}=\exp (|y|^{-z})$
indicates. Comparable results would be expected to hold for the two other
routes to chaos \cite{schuster1}, intermittency and quasiperidicity, in
low-dimensional maps as these exhibit both $q$-phase transitions \cite{mori1}
and $q$-sensitivity to initial conditions \cite{tsallis3}, \cite{robledo4}.
Clearly, our results establish a new feasible numerical scheme to determine
the family of values for the entropic index $q$ associated to a critical
multifractal attractor (here via the diameter function $\sigma $).
Conversely, the computation of the sensitivity $\xi _{t}$ at the onset of
chaos offers an original means to evaluate Feigenbaum's trajectory scaling
function $\sigma $ or its equivalent for other critical attractors.
Additionally and remarkably, the perturbation with noise of this attractor
brings out the main features of glassy dynamics in thermal systems \cite%
{robledo5}.

\textbf{Acknowledgments}. We acknowledge support from CONACyT and
DGAPA-UNAM, Mexican agencies.


\begin{references}
\bibitem{note1} This is not a generalized Pesin identity that would involve a generalized
 Kolmogorov-Sinai entropy. We discuss this issue in a subsequent Section..
\end{references}
\end{document}